# Icosahedral quasicrystal enhanced nucleation in commercially pure Ni processed by selective laser melting

C. Galera-Rueda[1,2], X. Jin[1,*], J. LLorca[1,2], M.T. Pérez-Prado[1,*]

[1] *Imdea Materials Institute, Calle Eric Kandel, 2, 28906 Getafe, Madrid, Spain.*

[2] *Department of Materials Science, Polytechnic University of Madrid/Universidad Politécnica de Madrid,*

*E. T. S. de Ingenieros de Caminos, Madrid 28040, Spain.*

*Corresponding author

**Abstract**

This work provides unambiguous evidence for the occurrence of icosahedral quasicrystal (iQC) enhanced nucleation during selective laser melting of gas atomized commercially-pure Ni powders. This solidification mechanism, which has only been recently reported in a few alloys and has to date never been observed in pure metals, consists on the solidification of grains of the primary phase on the facets of iQCs formed due to the presence of icosahedral short range order in the liquid. The occurrence of iQC enhanced nucleation has been inferred from the observation in the SLM processed pure Ni samples of an excess fraction of partially incoherent twin boundaries and of clusters of twinned grain pairs sharing common <110> five-fold symmetry axes. This work further evidences that additive manufacturing methods may constitute an invaluable tool for investigating the fundamentals of solidification and for the design of unprecedented grain boundary networks.

Selective laser melting (SLM), currently a widespread additive manufacturing method [1-5], consists on the layer-by-layer fabrication of metallic samples by the rapid melting and solidification of powders using a laser beam and following a predefined three-dimensional model. This technique has dramatically widened the scope of earlier metal processing techniques providing a large freedom of design for the fabrication of multi-material components and for the generation of unprecedented microstructures [1-4].

SLM involves several steps taking place far from thermodynamic equilibrium. More specifically, the raw materials are spherically-shaped gas atomized powders [6] produced typically by the fast cooling of metallic liquid droplets in a low pressure environment at cooling rates of the order of $10^5$-$10^6$ K/s [7]. Additionally, solidification of the laser melted powders into a predefined component during processing takes place at cooling rates ranging approximately from $10^5$ to $10^7$ K/s [8]. It seems logical that these complex out-of-equilibrium conditions involve solidification mechanisms that differ from those associated to conventional thermomechanical processing methods such as casting or wrought forming. Such mechanisms, as well as the corresponding unprecedented microstructures that might be thus generated, remain largely unknown.

Most SLM activities, both fundamental and applied, have targeted metallic materials with applications within the aerospace and biomedical industries, including Ni superalloys [9], Al alloys [10], Ti alloys [11], and steels [12]. SLM has been particularly successful in Ni superalloys, which are endowed with extraordinary mechanical behavior and oxidation resistance at high temperatures and are widely used in gas turbine components [13,14]. Significantly less research has focused on commercially pure metals, which generally have more reduced commercial interest, but which could greatly facilitate fundamental studies. In particular, SLM of pure Ni powders has only been reported in three papers, to the authors' knowledge [15-17]. These works have analyzed the phenomenon of balling [15] and have reported parameter optimization studies to process fully dense samples [17]. The corresponding processing-microstructure-property relationships during SLM of pure Ni remain thus widely unexplored.

The aim of this work is to gain a deeper understanding of the solidification mechanisms during selective laser melting of commercially-pure nickel, which has been chosen as a model face centred cubic (fcc) metal. With that purpose, SLM is carried out under several processing conditions and the resulting microstructures are thoroughly characterized by electron microscopy techniques. The raw material utilized in this study is gas atomized pure Ni powder, purchased from TLS Technik. The composition is summaryized in Table 1. As illustrated in Fig. 1a, powder particles have a relatively spherical morphology, albeit with rough surfaces which are populated by satellites. The particle size distribution (Fig. 1b) is unimodal, with $d_{10}$=30.2 μm; $d_{50}$=42.0 μm; $d_{90}$=58.5 μm. The powder flowability, apparent density and tap density of the powder are, respectively, 15.29 s, 4.34 g/cm$^3$, and 4.90 g/cm$^3$.

Prisms with dimensions 4x7x7 mm$^3$ were manufactured via SLM in a Renishaw AM400 system. SLM was performed using a bidirectional scanning strategy with 67° rotation per layer, a layer thickness of 30 μm and a hatch distance of 90 μm. Process parameter optimization was carried out using laser power values comprised between 200 and 300 W and scan speed values ranging from 0.75 m/s to 1.2 m/s. The three combinations of laser power and scan speed selected for this study are listed in Table 2. They have been labelled from A to C, and are listed in descending order of volumetric energy density ($E_V$). $E_V$ is defined as $E_V = P/vth$, where $P$ is the laser power (in J s$^{-1}$), $v$ the laser scanning velocity (in mm s$^{-1}$), $t$ the layer thickness (in mm) and $h$ the hatch distance (in mm). $E_V$ ranges from 148 J/mm$^3$ in sample A to 92 J/mm$^3$ in sample C. In general, higher $E_V$ involves higher amount of heat transferred to the melted material and enhances the partial remelting of the previously deposited layers. Low $E_V$ values are associated with higher thermal gradients, faster solidification rates, and less intense heating of previously deposited layers. Table 1 also summarizes the density of the as-built prisms processed under the three conditions. The density was measured by image analysis from collages of optical micrographs of cross sectional areas covering the entire specimens. It can be seen that the prisms processed using condition C present relatively low density values (93%) due to insufficient melting, while the A and B samples exhibit much higher densities (99.27 and 99.67%, respectively).

The microstructure, the microtexture, and the nature of grain boundaries (GBs) in the samples were characterized by electron backscattered diffraction (EBSD) using a FEI Helios NanoLab DualBeam 600i focused ion beam field emission gun scanning electron microscope (FIB-FEGSEM) with a step size of 1.1 μm, an accelerating voltage of 20 kV, and a beam current of 2.7 nA. The grain size was calculated from the EBSD inverse pole figure maps taking into consideration boundaries with misorientations higher than 15º, in agreement with common standards for microstructure quantification. A more detailed characterization of selected grain boundaries was carried out by transmission electron microscopy (TEM) in a FEI Talos F200x microscope operating at 200 KV. TEM lamellae were milled with a FIB from the areas of interest using a trenching-and-lift-out method, as described in [18].

Figures 2 (a-c) illustrate the inverse pole figure (IPF) EBSD maps in the building direction (BD) corresponding to sections of pure Ni prisms parallel to BD for processing conditions A (Fig. 2a, $E_V$ = 148 J/mm³), B (Fig. 2b, $E_V$ = 99 J/mm³), and C (Fig. 2c, $E_V$ = 92 J/mm³). The grain morphology is irregular, as is typical of SLM processed metals [1] and, as shown in the inverse stereographic triangles included as insets in Figs. 2 (a-c), the texture is in all cases very weak. The grain area cumulative distribution (Fig. 2d), measured from the corresponding EBSD IPF maps, shows that the grain size coarsens with increasing $E_V$.

The misorientation distribution histograms corresponding to the prisms processed using the three SLM parameter sets investigated are compared in Fig. 3a with that corresponding to a random distribution of cubes, named MacKenzie distribution [19], which is characteristic of cast metals and alloys. An excess fraction of GBs with misorientations lower than about 7º is observed in all cases [20], a common result in metals processed by SLM [1]. Additionally, a second peak corresponding to boundaries with misorientations close to 60º, which is absent in the classic MacKenzie distribution, can be distinguished. This peak (see inset in Fig. 3a) reveals an excess fraction of twin boundaries in all printed samples. As shown in Fig. 3b, the height of the peak, which scales with the twin boundary fraction, is inversely proportional to the average grain size. The EBSD GB maps corresponding to the areas of the printed pure Ni cubes depicted in Figs. 2(a-c) are shown in Figs.3(c-e). Twin boundaries have been plotted as red lines, while the remaining high angle boundaries, with misorientation angles higher than 15º, are plotted as black lines. These maps confirm the presence of an excess fraction of twin boundaries whose morphology is significantly more rugged than that normally observed in deformation or annealing twin boundaries [21].

Similar observations of an anomalously high fraction of twin boundaries were reported recently in a few fcc alloys processed by conventional casting, namely Al–20 wt.%Zn with 1000 ppm of Cr [22], commercially pure Al with Ti additions [23], Au–28.4 at.%Cu–16.7 at.%Ag (yellow gold) [24] and Au–28.4 at.%Cu–16.7 at.%Ag (pink gold) with Ir additions [25], as well as in two alloys processed by SLM (Inconel 718 [26] and Al7075 doped with ZrH$_2$ microparticles [20]). The origin of the excess twin boundaries in all these cases was attributed to icosahedral quasicrystal (iQC) enhanced nucleation. This solidification mechanism, which was recently discovered, consists in the nucleation of the primary fcc phase (α) on top of the triangular facets of iQC nucleant particles formed as a consequence of icosahedral short-range order (ISRO) in the undercooled liquid [27]. A schematic illustrating the orientation relationship between the iQCs and the growing α crystals [27,28] is included as Supplementary figure 1. A regular icosahedron exhibits 20 equilateral triangular facets, 12 vertices and 30 edges. The axes linking opposite vertices correspond to five-fold symmetry axes, with a rotation angle of 72° (five rotations of 72° make 360°) [28]. As explained in detail in [28], solidification of the α phase takes place in such a way that {111}$_α$ planes grow parallel to the triangular facets of the icosahedron and <110>$_α$ directions grow parallel to the icosahedron edges and, thus, to the five-fold symmetry axes [278. Therefore, adjacent α grains growing on five iQC facets sharing a common <110> five-fold axis must be twin related (Supplementary figure 1), except for the fact that the 70.5°<110> misorientation (that is characteristic of coherent Σ3 boundaries in fcc lattices) is 72°<110> in the twinned α grains formed from the iQC templates. This 1.5° angular difference has to be accommodated at the boundary but the actual accommodation mechanism has not been reported to date [28].

An unequivocal proof of the occurrence of iQc enhanced nucleation is thus the simultaneous presence of an excess of twin boundaries and of clusters of twinned grains sharing a <110> five-fold symmetry axis [20,22,24,26-27]. Such clusters were indeed observed in the three pure Ni samples processed using SLM parameter sets A to C and representative examples are shown in Fig.4 and in the Supplementary figures 2 and 3, respectively. The grains included in each cluster have been labelled using the corresponding numbers for the three processing conditions. The twin relationships between pairs of grains within each cluster are shown by means of (110) pole figures and the common five-fold <110> axes shared by several groups of these twinned grains are highlighted using white dotted lines in the corresponding (110) pole figures. Taking all these orientation relationships into account, the facets of regular icosahedrons have been numbered according to the corresponding α grains for each cluster. Several arrows in the icosahedron indicate the common <110> five-fold symmetry axes.

Altogether, the above observations provide evidence of the occurrence of icosahedral quasicrystal enhanced nucleation during SLM of the pure Ni gas atomized powders. TEM observations of the manufactured specimens did not provide direct evidence of the presence of the nucleant iQC particles, confirming that they are metastable in nature, in agreement with earlier studies [20,22,24,26-28]. The decrease in the fraction of excess twin boundaries with increasing grain size reported in Fig. 3b is consistent with the gradual disappearance of the initial solidification structure as a consequence of enhanced grain growth in the samples processed using higher energy density values.

As mentioned earlier, the twin boundaries formed by icosahedral quasicrystal enhanced nucleation must accommodate a misorientation angle of 72° around <110> directions, an angle that is 1.5° larger than that characteristic of coherent $\Sigma 3$ twin boundaries. A representative twin boundary of the pure Ni sample processed using the C parameter set was characterized by TEM to elucidate the accommodation mechanism. Fig. 5a shows a bright field (BF) image of the TEM lamella, where one twin boundary is marked by a red rectangle. This boundary is shown at larger magnification in Fig. 5b, and the selected area diffraction (SAD) patterns corresponding to the neighboring matrix (M) and twinned (T) regions along a common <110> zone axis are depicted in Fig. 5c. The rotation angle about the common <110> direction, measured from the SAD pattern, is 70.5°, i.e., the value associated to coherent Σ3 boundaries. Higher resolution TEM imaging (Fig. 5d) evidenced that the investigated boundary is formed by numerous steps, containing facets aligned with parallel {111} planes (i.e., coherent twin boundary segments). The formation of such steps is, thus, the mechanism by which the excess 1.5° misorientation is accommodated. Van Swygenhoven et al. [29] performed ab-initio modelling to determine the structure of high angle boundaries with 65° and 75° misorientations about <110> axes in pure Ni. These two misorientation angles lie within the Brandon criterion range [30] with respect to coherent twin boundaries. Their predictions confirm that the simulated boundaries consist of coherent twin segments parallel to {111} planes and steps between them, which are mostly disordered regions. Our results are in full agreement with these results.

In summary, this study reports the occurrence of iQC enhanced nucleation in pure Ni processed by selective laser melting. Evidence for this mechanism is given in the form of an excess fraction of twin boundaries and of clusters of twinned grain pairs with five-fold orientation symmetry. The structure of these twin boundaries consists of coherent segments parallel to {111} planes and steps in between to accommodate an extra 1.5° misorientation angle with respect to the fully coherent Σ3 twin boundary. We expect that, given the rapid solidification conditions inherent to selective laser melting, iQC enhanced nucleation might be found in other pure metals and alloys. This research also proves that SLM can be utilized to tailor the grain boundary network and, in particular, to tune the coherency of twin boundaries.

**Acknowledgments**

This work was partially supported by the MAT4.0-CM project funded by the Madrid region under program S2018/NMT-4381 and by the European Research Council (ERC) under the European Union's Horizon 2020 research and innovation programme (Advanced Grant VIRMETAL, grant agreement No. 669141). Additional support from the project PID2019-111285RB-I00, awarded by the Spanish Ministry of Science, Innovation and Universities, is also acknowledged. CG acknowledges support by the Spanish Ministry of Science, Education and Universities through the Fellowship FPU 18/01328.

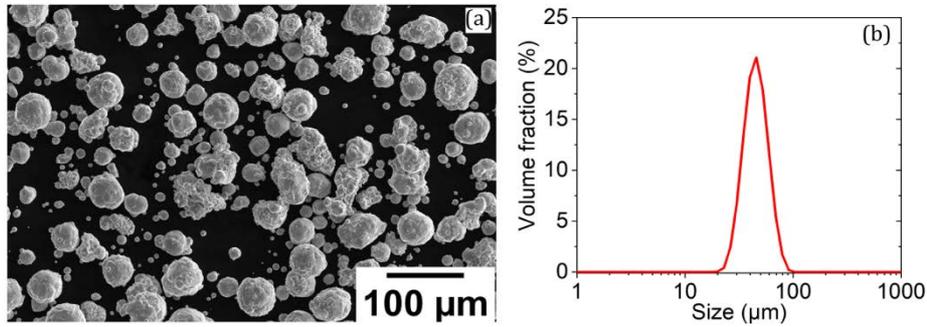

Figure 1. Commercially pure nickel powder utilized as raw material: (a) SEM secondary electron micrograph illustrating the powder morphology; (b) Particle size distribution.

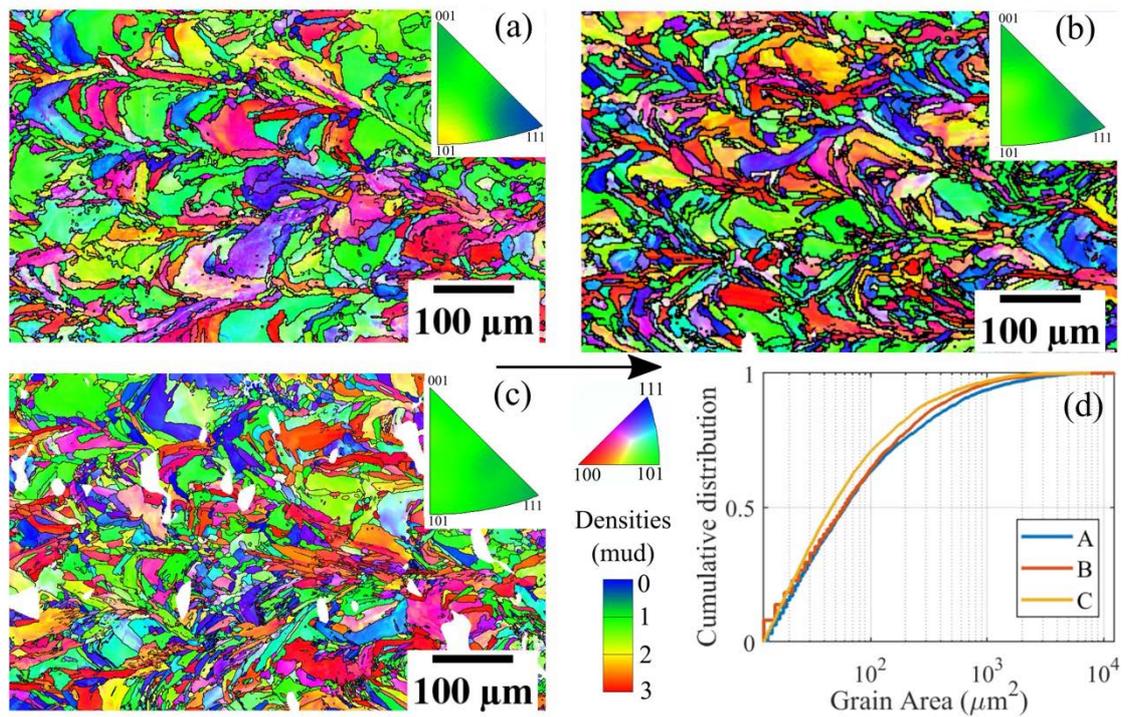

Figure. 2. EBSD inverse pole figure map in the building direction (BD is horizontal) and inverse pole figures illustrating the orientation of BD for the following processing conditions: (a) A, (b) B, (c) C; (d) Grain area cumulative distribution.

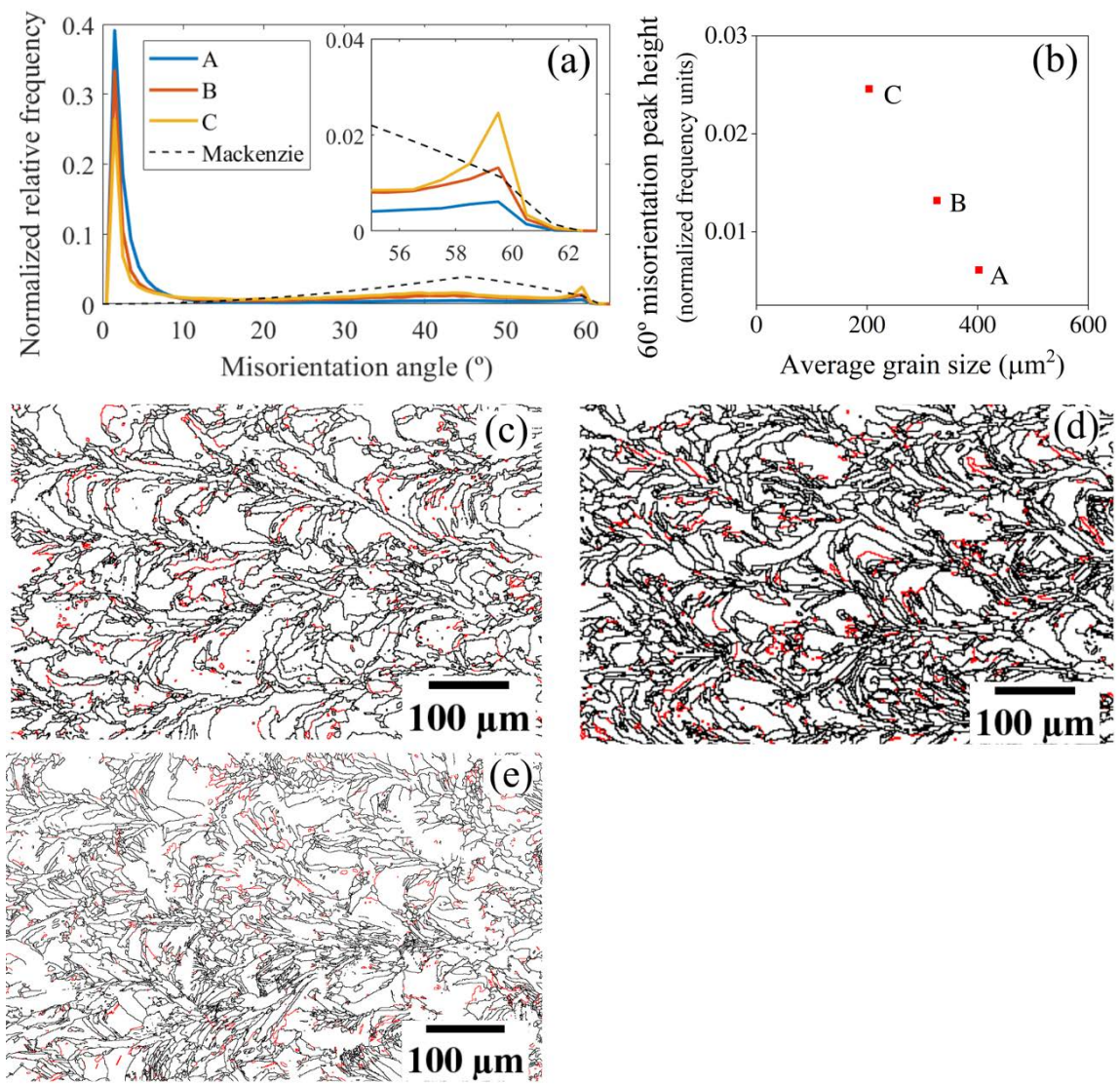

Figure 3. (a) Comparison of the grain boundary misorientation angle distributions corresponding to samples processed using conditions A to C with the MacKenzie distribution; (b) Height (in frequency units) of the 60° peak of the misorientation angle distribution with respect to the average grain size; (c-e) EBSD grain boundary maps corresponding to the specimens manufactured using the (c) A, (d) B, (e) C parameter sets. Twin boundaries are plotted in red and random high angle boundaries are plotted in black.

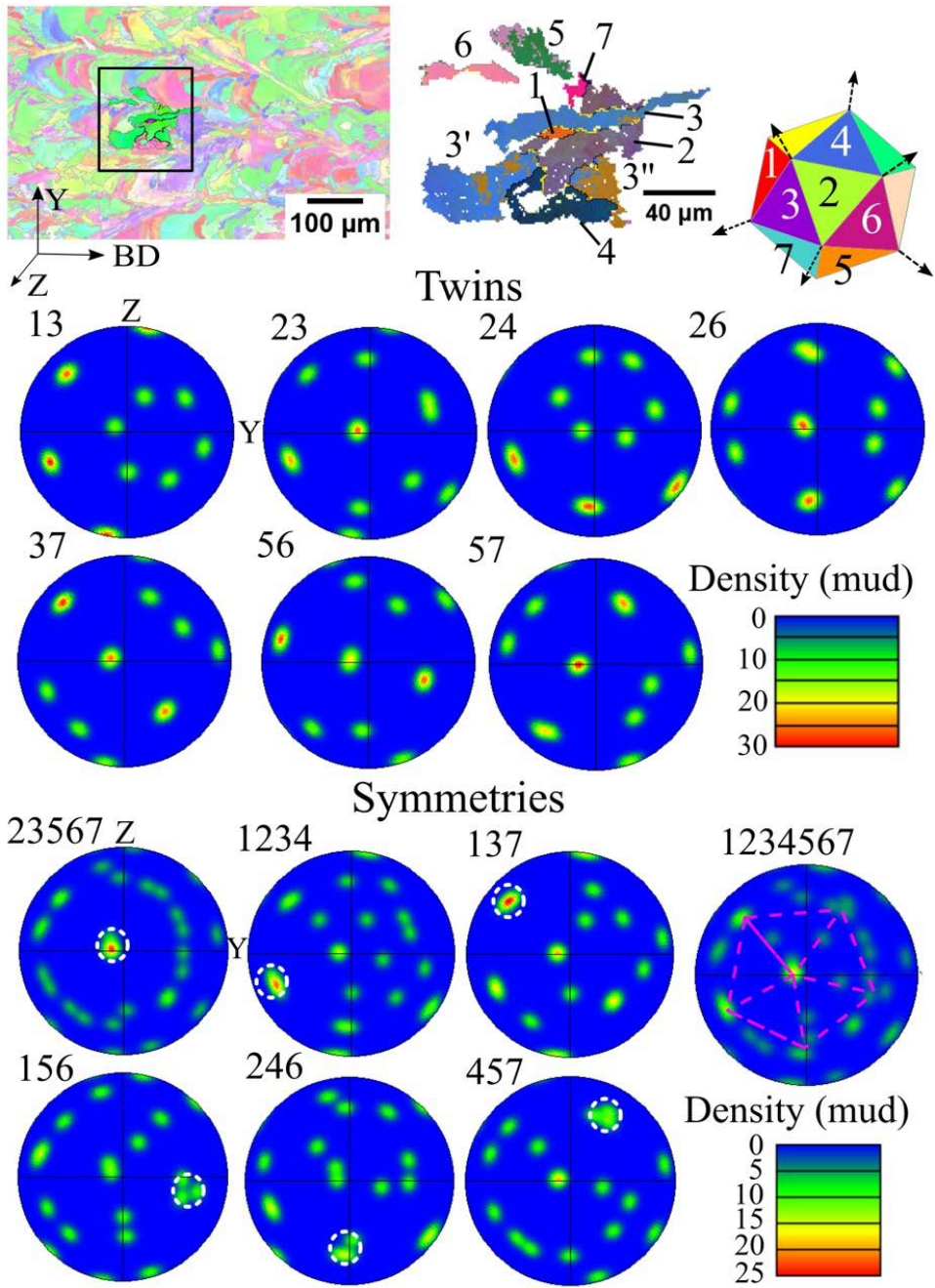

Figure 4. Example of one cluster of twinned grains sharing common five-fold <110> axes in the pure Ni prisms processed by SLM using condition A. The grains included in each cluster have been labelled using the corresponding numbers. The twin relationships between different grain pairs within each cluster are shown in (110) pole figures. The common <110> five-fold symmetry axes of different groups of grains are highlighted using white dotted lines also in (110) pole figures. An additional (110) pole figure, with larger size, compiles all the common <110> five-fold axes for each cluster, which are linked using dotted pink lines. Taking all the above orientation relationships into account, the facets of regular icosahedrons have been numbered according to the corresponding nucleated α grains. Several arrows in the icosahedron indicate the common <110> five-fold symmetry axes.

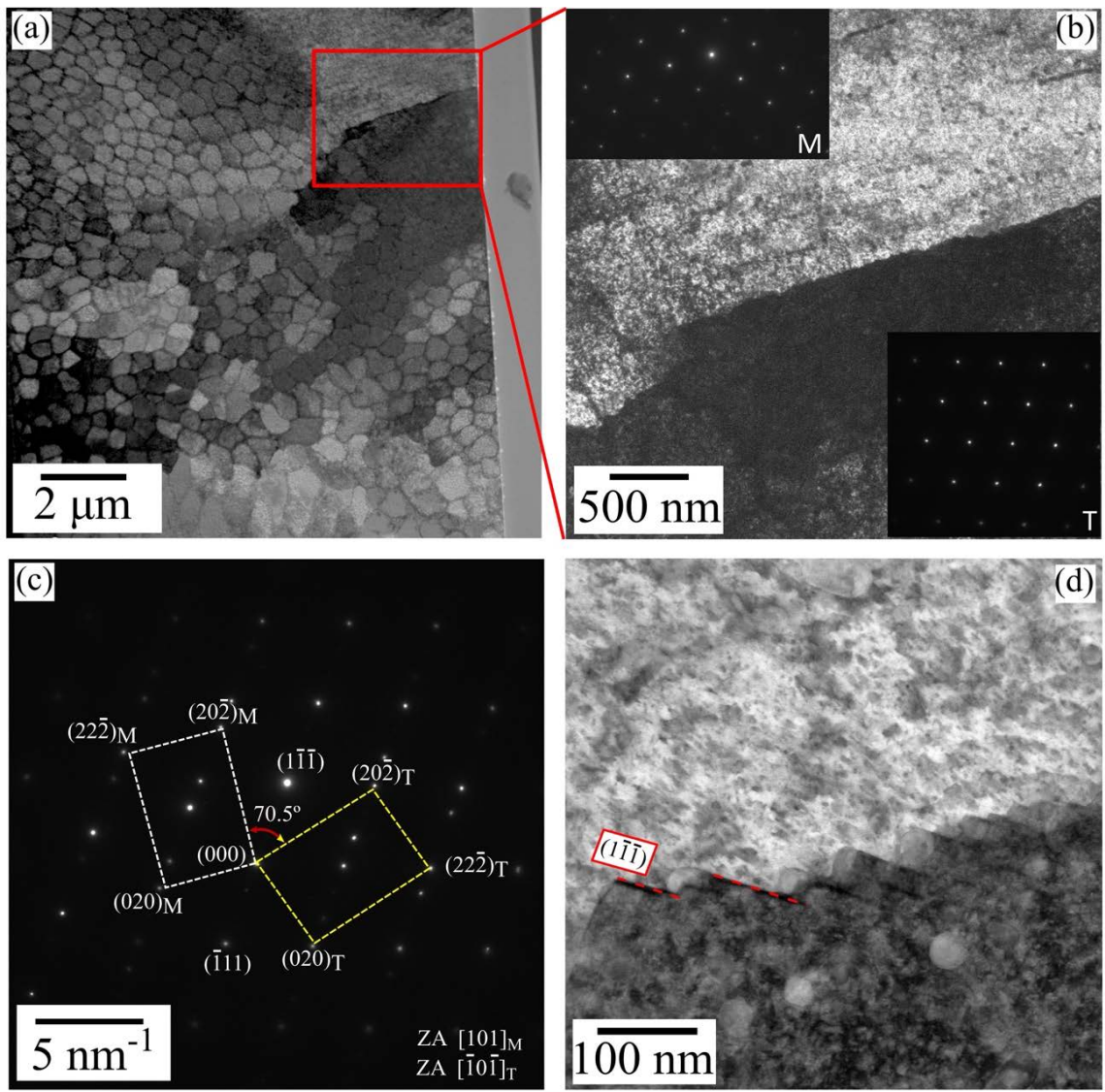

Figure 5. TEM examination of the structure of twin boundaries formed as a consequence of the icosahedral quasicrystal enhanced nucleation process. (a) Overview of the TEM lamella; (b) Enlarged view of one selected twin boundary; (c) SAD pattern illustrating the orientation relationship between the matrix (M) and the twin (T) located at both sides of the investigated boundary; (d) High magnification image of the faceted structure of the twin boundary. The coherent twin boundary segments, which are parallel to (111) planes, are highlighted in red.

15